\documentclass[conference]{IEEEtran}
\IEEEoverridecommandlockouts
\usepackage{yhmath}
\usepackage{booktabs}
\usepackage{caption}
\usepackage{subfigure}
\usepackage{multirow}
\usepackage{cite}
\usepackage{amsmath,amssymb,amsfonts}
\usepackage{algorithmic}
\usepackage{graphicx}
\usepackage{textcomp}
\usepackage{xcolor}
\def\BibTeX{{\rm B\kern-.05em{\sc i\kern-.025em b}\kern-.08em
    T\kern-.1667em\lower.7ex\hbox{E}\kern-.125emX}}
\begin{document}

\title{
SlimSpeech: Lightweight and Efficient Text-to-Speech with Slim Rectified Flow
\thanks{*Corresponding author}
 \thanks{This work was supported in part by the National Natural Science Foundation of China under Grants 62276220 and 62371407 and the Innovation of Policing Science and Technology, Fujian province (Grant number: 2024Y0068)}
}
\author{
	\IEEEauthorblockN{
		Kaidi Wang\IEEEauthorrefmark{2}, 
		Wenhao Guan\IEEEauthorrefmark{3}, 
		Shenghui Lu\IEEEauthorrefmark{2}, 
		Jianglong Yao\IEEEauthorrefmark{2}, 
		Lin Li\IEEEauthorrefmark{3}\IEEEauthorrefmark{1},
            Qingyang Hong\IEEEauthorrefmark{2}\IEEEauthorrefmark{1}
        } 
	\IEEEauthorblockA{\IEEEauthorrefmark{2}School of Informatics, Xiamen University, China}
	\IEEEauthorblockA{\IEEEauthorrefmark{3}School of Electronic Science and Engineering, Xiamen University, China 
    \\ kaidi@stu.xmu.edu.cn}
} 

\maketitle

\begin{abstract}
Recently, flow matching based speech synthesis has significantly enhanced the quality of synthesized speech while reducing the number of inference steps. In this paper, we introduce SlimSpeech, a lightweight and efficient speech synthesis system based on rectified flow. We have built upon the existing speech synthesis method utilizing the rectified flow model, modifying its structure to reduce parameters and serve as a teacher model. By refining the reflow operation, we directly derive a smaller model with a more straight sampling trajectory from the larger model, while utilizing distillation techniques to further enhance the model performance. Experimental results demonstrate that our proposed method, with significantly reduced model parameters, achieves comparable performance to larger models through one-step sampling.
\end{abstract}


\section{Introduction}
The objective of speech synthesis is to convert text into intelligible and natural speech. In recent years, neural network-based speech synthesis systems \cite{ren2020fastspeech, shen2018natural, li2019neural, li2021light} have made remarkable progress, significantly enhancing the quality and naturalness of synthesized speech. Most of these systems adopt a two-stage generation approach: first, an acoustic model converts text into acoustic features, and then a vocoder generates speech waveforms from these features. A significant portion of research on speech synthesis models focuses on the first stage, which plays a crucial role in determining the quality of synthesized speech. Diffusion probabilistic models (DPMs)\cite{ho2020denoising} have been widely applied in image and audio generation\cite{dhariwal2021diffusion,kong2020diffwave,jeong2021diff}. Acoustic models based on diffusion models\cite{guan23_interspeech} are capable of generating high-quality acoustic features, thereby advancing the field of speech synthesis.

However, DPMs require a substantial number of sampling steps during inference to produce a high-quality sample, which significantly limits the speed of speech synthesis, increases inference latency, and restricts the practical deployment of such models on edge devices. The challenge of ensuring high-quality output while reducing the number of inference steps has been a focal point of research\cite{song2020denoising} in recent years. ProDiff\cite{huang2022prodiff} proposes the use of a progressive distillation technique to reduce the number of inference steps. LightGrad\cite{chen2023lightgrad} accelerates the sampling process by leveraging DPM-Solver to derive the solution of the probability flow ordinary differential equation (ODE). DiffGAN-TTS\cite{liu2022diffgan} achieves high-fidelity and efficient text-to-speech (TTS) based on a denoising diffusion GAN model\cite{xiao2021tackling}. ComoSpeech\cite{ye2023comospeech}, on the other hand, introduces a consistency model\cite{song2023consistency} combined with distillation and utilizes one-step sampling to achieve satisfactory audio quality.

Recently, a novel generative model known as flow matching\cite{lipman2022flow,liu2022flow,guo2024voiceflow,guan2024lafma} has emerged, which directly learns an ODEs transformation from a standard Gaussian distribution to the real data distribution. Compared to diffusion models, it ensures the generation quality with a simpler approach and fewer steps. VoiceBox\cite{le2024voicebox} is the first one to employ the flow matching method to perform text-guided speech infilling tasks on large-scale training data. Matcha-TTS\cite{mehta2024matcha} directly utilizes optimal-transport conditional flow matching (OT-CFM) to train a TTS model. ReFlow-TTS\cite{guan2024reflow}, on the other hand, achieves high-fidelity speech synthesis through one-step sampling based on the Rectified Flow model. Despite its capability of one-step generation, it overlooks the size of model parameters.

\begin{figure*}[htbp]
    \centering
    \includegraphics[width=1\linewidth]{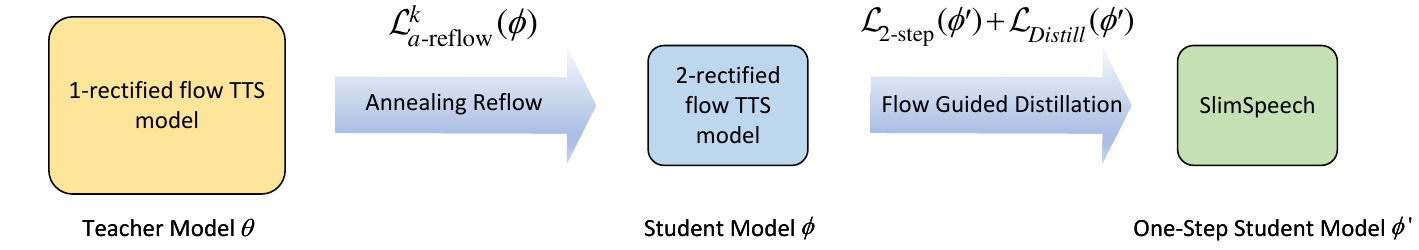}
    \caption{The training process of our proposed method.}
    \label{fig:fig1}
\end{figure*}

In this work, we explore the use of the rectified flow framework to jointly compress the parameter size and inference steps of the speech synthesis model. Inspired by slimflow\cite{zhu2024slimflow}, we introduce annealing reflow, which straightens the sampling trajectory under varying parameters, enhancing the sampling efficiency. Additionally, flow-guided distillation techniques are integrated to improve the quality of synthesized samples. Furthermore, depthwise separable convolutions are incorporated into the encoder to further minimize its parameters. The contributions of our work are as follows:

\begin{itemize}
\item We present SlimSpeech, a lightweight and efficient speech synthesis system leveraging the rectified flow model. Specifically, we propose to utilize annealing reflow in the speech synthesis model, which directly performs reflow operations from a larger, teacher model to obtain a smaller, student model, thereby avoiding initialization mismatch issues. Speech synthesis performance at fewer steps is further enhanced through distillation. Depthwise separable convolutions are employed to reduce the parameters of the text encoder.
\item Experimental results demonstrate that our model achieves comparable synthesis performance to larger models while significantly reducing the parameters, utilizing only one sampling steps.
\end{itemize}

\section{Background on rectified flow model}

In generative modeling, we aim to discover a mapping from a prior distribution to a data distribution. The rectified flow model\cite{liu2022flow} proposes leveraging an ordinary differential equation (ODE) to construct a continuous dynamical system that follows as straight a path as possible to generate the desired data distribution, requiring only a single step of computation to directly produce high-quality results. Specifically, given an initial prior distribution \(\pi _1\) and a target data distribution \(\pi _0\), we have the ODE:
\begin{equation}
dx_{t} = v_{\theta}(x_{t}, t)dt
  \label{equation:ODE}
\end{equation}
where \(t\in(0,1)\), and \( v_{\theta}\) denotes the vector field.
The rectified flow utilizes the following objective to train a vector field parameterized by a neural network \(\theta\):
\begin{equation}
\mathcal{L}_{rf}(\theta) = \mathbb{E}_{\mathbf x_{1}\sim \pi_1,\mathbf x_0 \sim \pi_0} 	\left[\int_{0}^{1}\lvert\lvert v_{\theta}(\mathbf x_t, t) - (\mathbf x_1 - \mathbf x_0)  \rvert\rvert ^{2} dt\right]
\end{equation}
where \(\mathbf x_t = t\mathbf x_1 + (1-t)\mathbf x_0\).

\subsection{Reflow}
To achieve a more direct probabilistic flow and fulfill the goal of one-step generation, given that the trajectory of the aforementioned ODE model may still be curved, the rectified flow introduces the reflow method to further straighten the ODE trajectory:
\begin{equation}
\mathcal{L}_{Reflow}(\phi) = \mathbb{E}_{\mathbf x_{1}\sim \pi_1} 	\left[\int_{0}^{1}\lvert\lvert v_{\phi}(\mathbf x_t, t) - (\mathbf x_1 - \mathbf{\hat x_0})  \rvert\rvert ^{2} dt\right]
\end{equation}
where \(\mathbf{\hat x_{0}}\) represents the data generated from the initial noise \(\mathbf x_{1}\) using the pre-trained probabilistic flow model \(v_{\theta}\) through the ODE in the equation (\ref{equation:ODE}). By continuing to train using the data from the ODE trajectory of \(v_{\theta}\) (named as 1-rectified flow), we obtain \(v_{\phi}\) (named as 2-rectified flow) which exhibits a more straight ODE trajectory, thereby enhancing sampling efficiency.

\subsection{Distillation}
The rectified flow framework also proposes utilizing distillation to enhance the effect of one-step generation:
\begin{equation}
\mathcal{L}_{Distill}(\phi') = \mathbb{E}_{\mathbf x_{1}\sim \pi_1} 	\left[ \mathbb{D} (ODE[v_{\phi}](\mathbf x_1), v_{\phi'}(\mathbf x_1, 1)) \right]
\end{equation}
where \(\mathbb D(\cdot,\cdot)\) represents the function for calculating the difference.
Besides, it is noteworthy that reflow and distillation can be used in combination: first, a more direct probabilistic flow model is obtained through reflow to generate better data pairs, which are then used for distillation. This combined approach has proven to be effective\cite{liu2022flow}.

\section{METHODOLOGY}
In this section, we provide a detailed explanation of our method. The training process is shown in the Fig \ref{fig:fig1}.
\subsection{Rectified Flow based Teacher Model}

Firstly, we train a large teacher model based on the rectified flow model as 1-rectified flow. Specifically, we train a parameter-reduced version of ReFlow-TTS model\cite{guan2024reflow}, which comprises four components: text encoder, duration predictor, length regulator, and rectified flow decoder. The structure of the duration predictor and length regulator remains consistent with FastSpeech2\cite{ren2020fastspeech}. For the text encoder, we introduce depthwise separable convolutions\cite{chollet2017xception} and set the channel dimension to 224 for lightweight purposes. The rectified flow decoder employs an architecture similar to DiffWave\cite{kong2020diffwave}, consisting of 20 stacked residual blocks with a channel dimension of 256. It utilizes sinusoidal position embedding\cite{vaswani2017attention} to obtain step embedding. 

Assuming that \(\pi _1\) represents the standard Gaussian distribution, and \(\pi _0\)  represents the true distribution of Mel-spectrogram data,
The training loss for the teacher model is:
\begin{equation}
\mathcal{L}_{rf}(\theta) = \mathbb{E}_{\mathbf x_{1}\sim \pi_1,\mathbf x_0 \sim \pi_0} 	\left[\int_{0}^{1}\lvert\lvert \mathbf v_{\theta}(\mathbf x_t, t, c) - (\mathbf x_1 - \mathbf x_0)  \rvert\rvert ^{2} dt\right]
\end{equation}
\begin{equation}
\mathcal{L}_{all}(\theta) = \mathcal{L}_{rf}(\theta) + \mathcal{L}_{dur}(\theta)
\end{equation}
where c represents text embeding.

\subsection{SlimFlow for TTS}\label{AA}
We propose to train a one-step text-to-speech student model using SlimFlow\cite{zhu2024slimflow} which incorporates annealing reflow and flow-guided distillation. Specifically, instead of training the entire model \cite{guo2024voiceflow,guan2024reflow}, we directly train a decoder with smaller parameters, while keeping the other modules from the teacher model and freezing their parameters.

\subsubsection{Anealing reflow}
Although the reflow stage can train a probabilistic flow with a straighter sampling trajectory, thereby reducing sampling steps and enhancing efficiency, it does not consider reducing the number of model parameters. We propose utilizing Annealing Reflow to directly train a smaller student model with an even straighter trajectory, overcoming the issue of parameter mismatch between the initialization of the teacher and student models. This approach smoothly transitions from training a 1-rectified flow to a 2-rectified flow, accelerating the model training process. The objective of annealing reflow is defined as follows:

\begin{equation}
\begin{aligned}
    \mathcal{L}_{a\text{-reflow}}^k(\phi) = 
    \mathbb{E}_{\mathbf{x}_1, \mathbf{x}_1' \sim \pi_1} \Bigg[ \int_0^1 & \| \mathbf{v}_\phi ( \mathbf{x}_t^{\beta(k)}, t, c ) - \\
    &( \mathbf{x}_1^{\beta(k)} - \hat{\mathbf{x}}_0 ) \|_2^2 dt ],
    \\
    \end{aligned}
\end{equation}

\begin{equation}
    $$
    where
    $$
    \mathbf{x}_t^{\beta(k)} = (1 - t) \hat{\mathbf{x}}_0 + t \mathbf{x}_1^{\beta(k)},
    $$$$
    \mathbf{x}_1^{\beta(k)} = \left( \sqrt{1 - \beta^2(k)} \mathbf{x}_1 + \beta(k) \mathbf{x}_1' \right),
    $$$$
    \hat{\mathbf{x}}_0 = \text{ODE}[\mathbf{v}_\theta](\mathbf{x}_1) = \mathbf{x}_1 + \int_1^0 \mathbf{v}_\theta(\mathbf{x}_t, t, c) dt.\nonumber
\end{equation}
\setcounter{equation}{7}
In the equation, \(k\) represents the number of training iterations, where \((\mathbf{x_1},\mathbf{\hat x_0},c)\) denotes the data pairs generated by the pre-trained teacher model. We define \(\beta (k)\) as follows:
\begin{equation}
    \beta (k) = 1 - min(1, k/K_{a-step})
\end{equation}
where \(K_{a-step}\) represents a constant.

It is noteworthy that as the training progresses, the training data gradually shifts from random data pairs to the data pairs generated by the pre-trained 1-rectified flow model, thereby ensuring the initialization of the student model and directly outputting a smaller 2-rectified flow model.

\begin{table*}[htb]
  \caption{Evaluation Results of Different Models. The RTF tests were conducted in both GPU
 (GeForce RTX 2080 Ti) and CPU (Intel(R) Xeon(R) CPU
 E5-2680 v4) environments, with the CPU tests performed
 on a single thread.}
  \label{tab:performence}
  \centering
  \begin{tabular}{lccccccc}
    \toprule
     Model & Sampling Steps ($\downarrow$) & FAD ($\downarrow$) & MOS ($\uparrow$) &  FD ($\downarrow$) & $\text{RTF}_{\text{gpu}}$ ($\downarrow$) & $\text{RTF}_{\text{cpu}}$ ($\downarrow$) & \#Params \\
    \midrule
    Ground truth (mel+vocoder)  & - & 0.303 & 4.41 ± 0.06 & 0.738 & - & - \\
    ReFlow-TTS (RK45 solver) & 179 & 0.335 & 4.21 ± 0.12 & 0.938 & 0.6503 & - & 27.09 M    \\
    Grad-TTS  & 4 & 0.457 & 3.64 ± 0.07 & 1.583 & 0.0186 & 
     0.9235 & 14.86 M \\
    Matcha-TTS & 4 & 0.950 & 3.89 ± 0.06 & 3.190 & 0.0153  & 
    0.1227 & 18.22 M \\
    2-ReFlow-TTS (Euler solver) & 4 & 0.338  & 4.04 ± 0.07 & 0.772 & 0.0133 & 0.1684 & 27.09 M   \\
    \midrule
    FastSpeech2 & 1 & 2.164 & 3.55 ± 0.08 & 6.025 & 0.0077 & 0.0759 & 28.83 M   \\    
    Grad-TTS  & 1  & 1.638 & 3.42 ± 0.08 & 2.771 & 0.0118 & 
    0.2441 & 14.86 M \\
    Matcha-TTS & 1 & 2.632 & 3.53 ± 0.08 & 9.714 & 0.0088 & 
    0.0354 & 18.22 M \\
    ReFlow-TTS (Euler solver) & 1 & 1.405 & 3.57 ± 0.06 & 4.257 & 0.0072 & 0.0477 & 27.09 M   \\
    2-ReFlow-TTS (Euler solver) & 1 & 0.486 & 3.76 ± 0.09  & 0.804 & 0.0074 & 0.0487 & 27.09 M   \\
    \midrule
    SlimFlow-TTS[teacher] (RK45 solver) & 164 & 0.349 & 4.18 ± 0.07 & 0.765 & 0.5394 & - & 17.61 M  \\
    SlimSpeech (Euler solver) & 4 & 0.674 & 3.94 ± 0.09 & 0.845 & 0.0130 & 0.0435 & 5.48 M \\
    SlimSpeech (Euler solver) & 1 & 0.693 & 3.71 ± 0.06 & 0.806 & 0.0080 & 0.0139 & 5.48 M \\
    \bottomrule
  \end{tabular}
\end{table*}

\subsubsection{Flow-Guided distillation}
Due to the limited capabilities of the student model, directly applying naive distillation may yield suboptimal results. To enhance the one-step generation capability of the student model while maintaining the dataset size, we employ flow-guided distillation. In addition to direct distillation, we introduce an additional 2-rectified flow based on few-step generation as a regularization term. Specifically, we obtain another two-step generation distillation loss:

\begin{equation}
\begin{aligned}
    \mathcal{L}_{2\text{-step}}(\phi') = &\mathbb{E}_{\mathbf{x}_1 \sim \pi_1} \Bigg [ \int_0^1 \mathcal{D}(\mathbf{x}_1 - (1 - t)\mathbf{v}_\phi(\mathbf{x}_1, 1, c) -\\&t\mathbf{v}_\phi(\mathbf{x}_t, t, c), \mathbf{x}_1 - \mathbf{v}_{\phi'}(\mathbf{x}_1, 1, c)) dt \Bigg ]
\end{aligned}
\end{equation}

where \(\mathcal{D}\) represents the L2 loss.

The total loss for this process is:
\begin{equation}
\mathcal{L}_{FG-Distill} = \mathcal{L}_{Distill}(\phi') + \mathcal{L}_{2\text{-step}}(\phi')
\end{equation}
\begin{equation}
\mathcal{L}_{Distill}(\phi') = \mathbb{E}_{\mathbf x_{1}\sim \pi_1} 	\left[ \| ODE[v_{\phi}](\mathbf x_1, c)- v_{\phi'}(\mathbf x_1, 1, c)\|^2 \right]
\end{equation}
\section{Experiments}
\subsection{Data}

We employ the LJSpeech dataset to evaluate our model, which comprises approximately 24 hours of female single-speaker audio recordings, totaling 13,100 samples. Out of these, we randomly select 100 samples as the validation set, 655 as the test set (5\%), leaving the remaining 12,345 samples for the training set. All audio recordings are converted into 80-dimensional Mel spectrograms, with a frame size and window size set to 1024 and a hop size of 256.

\subsection{Model Setup}
Firstly, we train a teacher model as 1-retified flow, which is a parameter-reduced version of ReFlow-TTS (named as SlimFlow-TTS). We employ the Adam optimizer to train the model for 240k iterations on two 2080ti GPUs. Upon completion of training, we utilize the RK45 solver to save data pairs \((\mathbf{x_1},\mathbf{\hat x_0},c)\). Subsequently, we directly utilize the saved data pairs to train a decoder with an even smaller parameter set (residual channels reduced from 256 to 96) for 240k iterations. The \(K_{a-step}\) in annealing reflow is set to 70k. Finally, by utilizing the RK45 Solver once again to generate new data pairs, we continue the distillation training for the decoder for an additional 160k iterations with one 2080ti GPU, yielding our final model.

We compare our model with FastSpeech2, Grad-TTS\footnote{https://github.com/huawei-noah/Speech-Backbones/tree/main/Grad-TTS}, Matcha-TTS\footnote{https://github.com/shivammehta25/Matcha-TTS}, and ReFlow-TTS. For the rectified flow based TTS model, we utilize the RK45 solver to generate high-fidelity spectrograms and employ the Euler solver for spectrograms generation with fewer steps. The obtained mel-spectrograms are converted into speech waveforms using a pre-trained HiFi-GAN \cite{kong2020hifi} model.

\subsection{Evaluation Metric}
We evaluated the performance of various systems, encompassing model parameters, Fréchet Audio Distance (FAD), Fréchet Distance (FD), Real-Time Factor (RTF), and subjective metric Mean Opinion Score (MOS) and comparative mean opinion score (CMOS). Model parameters directly mirror the model size. FAD and FD are metrics derived from the FID used in image generation, adapted to audio generation for evaluating the similarity between generated and real samples \cite{liu2023audioldm, ye2023comospeech, guan2024reflow}. In this paper, we adopt the implementation approach from \cite{liu2023audioldm}, where FAD utilizes the VGGish classifier for feature extraction, whereas FD employs the PANNs classifier for feature extraction. RTF reflects the model's ability to synthesize speech in real-time; for diffusion-based speech synthesis systems, a higher number of inference steps results in a higher RTF. Furthermore, we conducted subjective tests to evaluate the quality of generated speech. For each system, 15 audio samples were selected, and each audio was rated by 10 listeners on a scale of 1-5, with higher scores indicating better speech quality. We also choose CMOS (from -3 to 3) as a subjective metric to directly compare samples from the two systems.

\subsection{Main Result}

Table \ref{tab:performence} presents our experimental results. Notably, when compared to the original ReFlow-TTS, our teacher model, even with reduced parameters, exhibits negligible performance loss when tested with the RK45 solver, maintaining excellent performance. Therefore, the generated data accurately reflects the true data distribution, enabling effective training of the student model and inheriting the superior performance of the teacher model. Additionally, as shown in the Table \ref{tab:performence}, when the sampling step is set to 1, SlimSpeech achieves impressive results on FAD and FD, second only to 2-ReFlow-TTS, which contains approximately five times more parameters than SlimSpeech. Meanwhile, its MOS score is comparable to that of larger models. This suggests that our method demonstrates strong modeling capabilities for complex speech data while reducing the number of parameters.

Furthermore, we tested the multi-step generation performence. Specifically, after obtaining a model with a smaller number of parameters through the annealing reflow, we employed the distillation method to generate a 4-step model. As show in the Table \ref{tab:performence}, when increasing the inference steps, the SlimSpeech's MOS score surpasses mos t systems while maintaining low FAD and FD. This demonstrates that although our model can produce satisfactory results with a single inference step, we can also train a smaller model with better performance by increasing the inference steps.

We also evaluated the efficiency of different models by testing their RTF. Generally, the results shows that models with more parameters and higher sampling steps tend to have slower inference speeds. However, our model, despite having fewer parameters but the same number of sampling steps, achieved a similar inference speed to the larger model (ReFlow-TTS) on GPU. We attribute this to the performance advantages of GPUs. On the other hand, on the CPU, our model's inference speed was nearly four times faster than that of the larger model, demonstrating its suitability for resource-constrained environments.

\begin{table}[htb]
  \caption{Ablation Study}
  \label{tab:objective2}
  \centering
  \begin{tabular}{lcccc}
    \toprule
    Model & Sampling Steps & FAD & FD & CMOS \\
    \midrule
    2-rectified flow TTS & 96 & 0.607 & 0.810 & 0 \\
    w/o annealing reflow & 96 & 0.643 & 0.893 & -0.07 \\
    \midrule
    SlimSpeech & 1 & 0.693 & 0.806 & 0 \\
    w/o \(\mathcal{L}_{2\text{-step}}\) & 1 & 0.954 & 0.844 & -0.04 \\
    w/o \(\mathcal{L}_{2\text{-step}}\) + \(\mathcal{L}_{Distill}\) & 1 & 1.114 & 0.915 &  -0.47\\

    \bottomrule
  \end{tabular}
\end{table}

\subsection{Ablation Study}
We also employ ablation experiments to demonstrate the effectiveness of annealing reflow and flow-guided distillation. Fitstly, we directly train a smaller student model using data generated by the teacher model, employing the RK45 solver to produce samples. As seen in Table 2, adopting annealing reflow to train the student model yields superior performance. Becides, during the distillation phase, when we incorporate an additional 2-step distillation loss, FAD improves from 0.954 to 0.693, accompanied by a slight enhancement in FD and CMOS, underscoring the superiority of our approach. When the distillation operation is canceled, there is a significant decrease in performance on both objective and subjective metrics, demonstrating the importance of distillation.

\section{CONCLUSION}

In this paper, we propose SlimSpeech, a lightweight and efficient speech synthesis model using rectified flow. We refine the ReFlow-TTS model architecture to directly train a teacher model based on rectified flow. By utilizing SlimFlow, we further optimize the reflow and distillation operations within the rectified flow framework, enabling our model to achieve high efficiency while significantly reducing the number of parameters, with one-step generation performance comparable to that of larger models. Audio samples are available at https://wkd88.github.io/.

\section*{Acknowledgment}
We would like to thank Xinhua Song, Suxia Xu and Limin Lai at Xiamen University for their support in this work.

\bibliographystyle{IEEEbib}
\bibliography{strings}

\end{document}